\documentclass[12pt]{elsart}
\usepackage{amsfonts}
\begin{document}

\begin{frontmatter}
\title{Using network-flow techniques to solve an optimization problem from surface-physics}
\author[zpr]{U.~Blasum}
\author[zpr]{W.~Hochst{\"a}ttler}
\author[zpr]{C.~Moll}
\author[juelich]{H.~Rieger}
\address[zpr]{Zentrum f\"ur paralleles Rechnen, Universit\"at zu K\"oln, 50937 K\"oln, Germany}
\address[juelich]{HLRZ c/o Forschungszentrum J\"ulich, 52425 J\"ulich, Germany}
\begin{abstract}
The solid-on-solid model provides a commonly used framework
for the description of surfaces. In the last years it has been extended in order to investigate
the effect of defects in the bulk on the roughness of the surface. 
The determination of the ground state of this model leads to a combinatorial
problem, which is reduced to an uncapacitated, convex minimum-circulation problem. We will
show that the successive shortest path algorithm solves the problem in polynomial time.
\end{abstract}
\end{frontmatter}

\section{Introduction}
In the past twenty years there has been a fruitful collaboration 
between theoretical physicists and researchers from the field of 
combinatorial optimization on the determination of the global 
energy-minimum (ground state) in various spin-glass models. 
The observation, that the ground state of the Ising-model 
can be found by solving an certain maxcut-problem, has drawn 
the attention of physicists to the methods developed in graph theory 
\cite{bieche:groundstates}. Conversely 
this application has animated the research on the maxcut-problem 
\cite{barahona:branchcut,Juenger95} itself. 
It the same spirit it was shown, that in the context 
of the Random-Field-Ising-model the maxflow-problem arises \cite{barahona:rfim}. 

The {\em solid-on-solid (SOS) model}, on which we will comment here, is a 
microscopic approach to describe the aggregation of atoms on a crystalline 
substrate. In contrast to the above mentioned Ising-models, where the dynamic 
variables can take on only two values, in the SOS-model these variables are integer valued.
At first sight this fact seems to complicate the optimization problem. However, 
similar to the Ising-model, the underlying structure can be represented 
by a graph. We will show in this paper, that in the planar  case the problem of 
minimizing the surface-energy can be reduced to the problem of finding a 
network-circulation, that minimizes a convex objective function.

The growth of surfaces is a subject of intense research in chemistry and solid state 
physics. A deeper understanding of the relevant processes has applications in the 
fabrication of semiconductors and various other fields where the precise shape of the 
surface is crucial \cite{kineticsofordering}. In the simple approach taken
by the SOS-model 
the lattice structure of real crystals is taken into account by subdividing the 
surface into discrete sites (e.g.~unit-squares) on which particles, modeled by 
unit cubes, can aggregate \cite{bcf:sos,chuiweeks:rough}. There are several possible 
parametrizations of the model. The one which is commonly used for theoretical 
investigations yields the surface-energy

\begin {equation}
H(h)=\sum _{\langle ij \rangle}(h_i-h_j)^2. \label {eq:sosmodel} 
\end{equation}

Here $h_i\in\Zset$ is the height of the surface at site $i$ and $\langle ij \rangle$ 
indicates all pairs of nearest neighbor sites on e.g.~a square lattice, mimicking the 
crystalline structure. Obviously the ground state of this model is flat, i.e.~$h_i=const$ 
for all $i$. Thermal fluctuations lead to a roughening transition at some temperature 
above which the height-height correlation function increases logarithmically with the distance.

One assumption of this model is the flatness of the underlying substrate, i.e.~a 
bulk-ordered crystal. Due to the presence of quenched defects one should consider the 
intriguing question of what the effect of a disordered substrate on the thermodynamic 
(and non-equilibrium growth) properties of the surface might be \cite{Toner}. It has 
been suggested \cite{Tsai92} to introduce a disordered substrate into the model 
(\ref{eq:sosmodel}) by providing each of the height variables with a real valued 
offset $d_i\in\Rset$, which is generally chosen to be from $[-0.5,0.5]$:

\begin{equation}
h_i:=n_i+d_i\quad{\rm with}\quad n_i\in \Zset. \label{eq:height}
\end{equation}

This model has some new features not contained in the original one. 
The influence of the disorder tends to roughen the surface {\em also at low temperatures}. 
In particular a new phase has been predicted \cite{Tsai92,marinari:superrough}, where 
height-height correlations diverge more strongly with the distance than logarithmically. 
Since this is a low temperature phase the properties of the ground state of (\ref{eq:sosmodel}) 
with (\ref{eq:height}), which is no longer trivial due to the competition between the elastic 
term $(h_i-h_j)^2$ and the random offsets $d_i$, are of crucial interest.  

The paper is organized as follows: Section \ref{sec:reduction} introduces the reduction 
to a convex network-flow problem and in section \ref{sec:complexity} we will then show 
that (\ref{eq:sosmodel}) can be minimized in strictly polynomial time using the techniques 
introduced in \cite{Minoux86}. We use standard notation from graph theory, and refer 
the reader to \cite{Berge1985} for the definitions.

\section{Reduction to a network-flow problem\label{sec:reduction}}

Let $D=(V,E)$ be a directed graph. The optimization problem we consider is
\begin{eqnarray}
&\min H(n)=\sum _{(ij)\in E}((d_i+n_i)-(d_j+n_j))^2& \label {eq:nodeobjective} \\
&\mbox{s.t.}& \nonumber \\
&n_i\in \Zset \mbox{ forall } i\in V.& \nonumber
\end{eqnarray}

Using the definitions $d_{ij} :=d_i-d_j\mbox{ and   }x_{ij}:= n_i-n_j$
the objective-function can be written as
\begin {equation}
H(X)=\sum_{(ij)\in E}(d_{ij}+x_{ij})^2. \label{eq:edgeobjective} 
\end {equation}

We intend to reformulate the original optimization problem in terms of the variables 
$x_{ij}$. Obviously we must demand $x_{ij}\in \Zset$ for all edges $(ij)\in E$. Furthermore, since 
the $x_{ij}$ describe potential-differences in the scalar field given by a set of 
height variables $n_i$, it is clear that the sum of the $x_{ij}$ along any oriented 
cycle on the surface has to be zero. The following theorem establishes this constraint 
formally in the special case, where $D$ is a planar graph. 

\begin{thm}\label{theo:cocircuit}
Let $D=(V,E)$ be a planar, directed graph and $D^*=(V^*,E)$ be the associated dual
graph. There exists a set of node-variables $n_1, \ldots n_{|V|}$ such that
$X=(x_{ij})^E=(n_i-n_j)^E$ if and only if $X$ satisfies
\begin {equation}
\sum _{(ij)\in E}x_{ij}-\sum _{(ji)\in E}x_{ji}=0 \mbox{ for all dual nodes }i\in V^*. \label{eq:dualflow}
\end{equation}
\end{thm}

A proof of this theorem can be found in \cite{Berge1985} Chapter 5.

Given a vector $X$ and a node $i$ we call the number
\[g_i(X):=\sum _{(ij)\in E}x_{ij}-\sum _{(ji)\in E}x_{ji}\]
the excess of a node $i$. A complete formulation, which, for planar graphs, 
is equivalent to (\ref{eq:nodeobjective}) is thus given by
\begin {eqnarray}
&\min H(X)=\sum_{(ij)\in E}(d_{ij}+x_{ij})^2,& \label{eq:sos}\\
&\mbox{s.t.}& \nonumber \\
&x_{ij}\in \Zset \mbox{\hspace{1cm}for all edges } (ij)\in E,&  \label{eq:sosint}\\
&g_i(X)=0  \mbox{\hspace{1cm}for all dual nodes }i\in V^*.& \label{eq:sosflow}
\end{eqnarray}
A vector $X$ satisfying these equations is called a circulation \cite{ahuja:netflow}. 
The problem is a convex network-flow problem in the dual graph without any capacity constraints 
(the flow-variables $x_{ij}$ on each edge can take on arbitrary positive or negative values). 
We note that this reduction, as well as the results in the following section, more generally 
hold for any objective function, which can be written as 
\[H(X)=\sum _{(ij)\in E} \theta _{ij}(x_{ij})\]
where the $\theta _{ij}$ are convex functions with a minimum.

\section{Algorithm and complexity \label{sec:complexity}}

The algorithms which solve the convex flow problem \cite{Minoux86}
are not strictly polynomial if there are no capacity constraints. 
We will show in this section that in the special case of  problem (\ref{eq:sos}) 
the time complexity of the successive shortest path algorithm for convex flow can 
be bounded by a strict polynomial. For this purpose 
we start with a brief review of the algorithm, a more detailed description can be found 
in \cite{ahuja:netflow}. 

The underlying philosophy is to search the solution in a subset of $\Zset^E$, in which each 
vector $X$ yields a lower bound for the optimum of (\ref{eq:sos}) but does not necessarily 
satisfy the flow-constraints (\ref{eq:sosflow}). The flow balance at all nodes has to be 
approached iteratively by the algorithm. To characterize this subset, one introduces 
the notion of dual feasibility:

\begin{defn}
Let $H(X)=\sum_{(ij)\in E}\theta_{ij}(x_{ij})$ be a sum of functions $\theta _{ij}(x)$
of the components $x_{ij}$ of $X$. A vector $X$ is called dual feasible, 
if there is a number $\pi_i$ for every node (potential-function), such that 
for all edges $(ij)\in E$ the reduced cost inequalities
\begin{eqnarray}
c_{ij}^{\pi}(x_{ij}):=\theta_{ij}(x_{ij}+1)-\theta_{ij}(x_{ij})-\pi_j+\pi_i\ge0 \label{eq:dualfeas}\\
c_{ji}^{\pi}(x_{ij}):=\theta_{ij}(x_{ij}-1)- \theta_{ij}(x_{ij})+\pi_j-\pi_i\ge0 \nonumber
\end{eqnarray}
are satisfied.
\end{defn}

\begin{thm}
Let $X$ be a dual feasible vector and let $X_{\mathrm{opt}}$ be an optimal solution
of problem (\ref{eq:sos}), then $H(X)\le H(X_{\mathrm{opt}})$. 
\end{thm}

The proof is given in \cite{ahuja:netflow} chapter 9 and chapter 14.3. 
An immediate consequence of this theorem is
\begin{cor}
A dual feasible vector $X$ satisfying (\ref{eq:sosint}), (\ref{eq:sosflow}) 
is an optimal solution of problem (\ref{eq:sos}).
\end{cor}

In order to apply the successive shortest path algorithm we determine
a dual feasible initial solution. Here we can choose the vector $X_s\in \Zset^E$, 
whose components minimize the functions $\theta_{ij}(x)$ and the potential function 
$\pi:=(\pi_i)^{|V|}=0$. 
Note that, when $\pi=0$, the reduced costs (\ref{eq:dualfeas}) simply measure the 
change of the objective function triggered by a change of the 
$(ij)$--component by plus resp. minus one. These are clearly all positive if $X$ 
minizes every summand of (\ref{eq:sos}).

The initial solution does in general not satisfy the flow-equations 
(\ref{eq:sosflow}). The excess $g_i(X)$ of all nodes is then balanced by 
iteratively augmenting $X$ along paths between supply-nodes ($g_i(X)>0$) and 
demand-nodes ($g_i(X)<0$), which are shortest paths with respect to the reduced costs.
The variables $x_{ij}$ associated with the edges along such a path are changed by exactly one unit (the sign depends on the orientation of the edge in the path). 

Due to the non-linearity of the objective every update of $X$ leads to a new set 
of arc-weights. However, the convexity of the individual functions $\theta _{ij}$
ensures, that the inequalities (\ref{eq:dualfeas}) still hold after every
augmentation, when $c_{ij}^{\pi}(x_{ij})$ is replaced by
$c_{ij}^{\pi}(x_{ij}+a)$ with $a\in \{-1,0,1\}$. Therefore the algorithm 
works correctly because it always maintains dual 
feasibility of $X$ with respect to the current set of arc-weights. 

\begin{tabbing}
ww\=ww\=ww\=ww\=ww\=ww\=ww\=ww\=ww\=ww\kill 
\+ {\bf procedure} successive shortest path \\
\+ {\bf begin} \\
$X=\ min \{H(X)|X\in \Zset^E\}$ \\
$\pi=0$ \\
\+ {\bf while} there is a node $s$ with $g_s(X)>0$ \\
compute the reduced costs $c^{\pi}(X)$\\
\+determine the shortest path distances $d(i)$ from $s$ to all \\
\-other nodes with respect to the reduced costs \\
choose a node $t$ with $g_t(X)<0$ \\
let $w(s,t)$ denote the shortest path vector from $s$ to $t$ in $D$\\
\label{alg:xupdate} $X=X+w(s,t)$ \\
\-$\pi=\pi-d$ \\
\-{\bf end} \\
{\bf end} \\
\end{tabbing}
 
The complexity of the algorithm is $O(G(X_s))S(n,m)$. Here $S(n,m)$ is the 
time needed for a shortest path search in a network with positive arc-weights and
\[G(X_s):=\frac{1}{2}\sum _i |g_i(X_s)|\]
counts the number of paths needed to balance the excess of all nodes. 
The following lemma then yields an upper bound for the value of $G(X_s)$.

\begin{lem}
The vector $X_s\in \Zset^E$ that minimizes $H(X)$ 
(eq.~\ref{eq:sos}) satisfies $G(X_s)\le|E|$.
\end{lem}

\begin{pf}
If the integrality- as well as the flow--constraints are neglected the global 
minimum of $H(X)=\sum_{(ij)\in E}(d_{ij}+x_{ij})^2$ is attained at $-d:=(-d_{ij})^E$ 
with $H(-d)=0$. If we now take into account the integrality--constraints 
it follows from the convexity of the $(d_{ij}+x_{ij})^2$, that the minimum 
$x_{ij}^s\in \Zset$ is attained in either $\lceil -d_{ij} \rceil$ or 
$\lfloor -d_{ij} \rfloor$. We thus find 
\[|x_{ij}^s-(-d_{ij})|\le 1.\]
From theorem \ref{theo:cocircuit} we obtain $G(-d)=0$ and thus
\[G(X_s)=G(X_s)-G(-d)\le \sum_{(ij)\in E}|x_{ij}^s-(-d_{ij})|\le |E|.\]
\end{pf}

\section{Computer experiments}

We implemented the successive shortest path algorithm for the
convex-flow problem. The program needs
less than 80 seconds on a Sparc Workstation to solve instances of grid-graphs with
$128\times128$ nodes and randomly chosen offsets to optimality.
Grid-graphs of size $200\times200$ nodes take about 10-15 minutes. We
used C++ for our implementation and made use of the LEDA-class library
\cite{ledamanual}.  The physical results, which have been 
obtained with this program, together with their theoretical interpretation 
will be reported elsewhere \cite{unpub}.

\end {document}